\documentstyle{cupconf}

\ifoldfss
\else
  \ifnfssone
    \newmathalphabet{\mathit}
      \addtoversion{normal}{\mathit}{cmr}{m}{it}
      \addtoversion{bold}{\mathit}{cmr}{bx}{it}
    \newmathalphabet{\mathcal}
      \addtoversion{normal}{\mathcal}{cmsy}{m}{n}
    \else
    \ifnfsstwo
    \fi
  \fi
\fi

\def\nuc#1#2{\relax\ifmmode{}^{#1}{\protect\text{#2}}\else${}^{#1}$#2\fi}
\def\hexnumber#1{\ifcase#1 0\or1\or2\or3\or4\or5\or6\or7\or8\or9\or
 A\or B\or C\or D\or E\or F\fi }
\makeatletter
\ifx\CUP@mtlplain@loaded\undefined
\else
\fi
\makeatother
 \makeatletter
 \ifx\CUP@mtlplain@loaded\undefined
   \font\tenbmi=cmmib10 at 10pt
   \font\sevenbmi=cmmib10 at 7pt
   \font\fivebmi=cmmib10 at 5pt

   \newfam\bmifam
   \textfont\bmifam=\tenbmi
   \scriptfont\bmifam=\sevenbmi
   \scriptscriptfont\bmifam=\fivebmi
   
 \fi
 \makeatother
\ifnfsstwo

\fi
\ifnfssone

\fi
\ifoldfss

\fi

\mathchardef\varLambda="0103

\makeatletter
\ifx\CUP@mtlplain@loaded\undefined
\else
\fi
\makeatother
\makeatletter
\ifx\CUP@mtlplain@loaded\undefined
  \font\tenbms=cmbsy10
  \font\sevenbms=cmbsy10 at 7pt
  \font\fivebms=cmbsy10 at 5pt
  \newfam\bmsfam
  \textfont\bmsfam=\tenbms
  \scriptfont\bmsfam=\sevenbms
  \scriptscriptfont\bmsfam=\fivebms

  \edef\bsy@{\hexnumber\bmsfam}
  \mathchardef\bnabla="0\bsy@72
\fi
\makeatother
\def\etal{\mbox{\it et al.}}

\title[The Biggest Explosion Since the Big Bang]{Universality in SNIae and
the Phillips Relation}

\author[D. Arnett]%
{D\ls A\ls V\ls I\ls D\ns A\ls R\ls N\ls E\ls T\ls T$^1$}

\affiliation{$^1$Steward Observatory, University of Arizona,
Tucson, AZ 85721, USA}

\begin{document}
\ifnfssone
\else
  \ifnfsstwo
  \else
    \ifoldfss
      \let\mathcal\cal
      \let\mathrm\rm
      \let\mathsf\sf
    \fi
  \fi
\fi

\maketitle

\begin{abstract}
The use of supernovae of Type~Ia for the determination of
 accurate distances rests upon
the empirical Phillips relation, in which the brightest events are
the broadest in time. Implications of new data upon the homogeneity
of light curves under the operation of a stretch in time, 
of the parabolic luminosity increase at the earliest times,
and of the time from explosion to maximum light
are discussed. The early luminosity is in excellent agreement
with the predictions of \cite{arn82}, and the lack of prominent
higher modes of diffusion constrain progenitor and explosion models.
Difficulties with reproducing the observed rise
time are restricted to radiative transfer models
(e.g., \cite{hoef}), and probably
due to an overestimate of thermal photon escape due to
inadequate line lists. Because of the strong dependence of
luminosity on \nuc{56}{Ni} mass, some simple models can give
a Phillips relation of the correct sense.
\end{abstract}

\firstsection 
\section{Introduction}

Supernovae of Type~Ia (SNIae)
 have been found at high redshift (up to $z \approx 1$)
\cite{perl97b,hiz,perl99a}. 
Being about $10^6$ times brighter than Cepheids, SNIae
can be seen about $10^3$ times further, and consequently provide
a more interesting possibility for determining truly cosmic distances.
Although the SNIae are {\em not} identical, so that each
event is not strictly like the standard
one, \cite{phillips} found that the brighter events are broader in
time (the Phillips relation). Calibration of this variation, 
with a set of relatively nearby supernovae, the Calan-Tololo sample
of \cite{calan}, allows 
individual events to be placed upon a standard scale
\cite{rpk}, and so doing reduces
the scatter in the Hubble diagram (a plot of redshift versus calibrated
brightness).

To properly use this breakthrough, we must understand the underlying
physics of Phillip's empirical relation. Several recent observational
developments,
\begin{enumerate}
\item The explicit demonstration, by  \cite{perl97a,perl.this},
 that the application of a time stretch
operation to SNIae and a luminosity normalization, give a universal
light curve shape, and

\item Measurement of the rise to maximum from low luminosities 
\cite{riess.near},
gives the detailed structure of the early light curve,
\end{enumerate}
provide interesting information on the nature of the supernova event.
The fact that SNIa light curves, except for a few odd events that can
be easily recognized by their spectra, represent a single parameter family
of shapes which can be ``stretched'' to a single universal shape, needs
to be explained. It is this universality that allows the nature of the
individual event to be identified, independent of distance, and their
use as distance indicators to be precise.


\section{Perspective on Analytic Solutions}

The luminosity of SNIae is provided almost entirely
by the decay of $\rm^{56}Ni$ and  $\rm^{56}Co$; other radioactive
sources become important well after maximum light, and thermal
emission from shocks is small.  
Determining the shape of the light curve, that is, the luminosity
as a function of time, involves three separate issues:
\begin{enumerate}
\item The transfer of radioactive energy by gamma-rays and positrons
to the thermal energy of the plasma, properly including energy 
which escapes in the form of positrons, gamma and x-rays,
\item The work done on the expanding plasma by the thermal radiation
(or alternatively, the reduction in radiation energy by the 
accumulated redshift of the trapped photons), and,
\item The energy escaping as thermalized radiation, which forms the
supernova spectrum observed from the infra-red through the
visible to ultra-violet (UVOIR).
\end{enumerate}
The analytic solutions used approximate the last item, the escape
probability, by a grey diffusion operator, parameterized by an
effective opacity. The actual physics of the escape of thermal
radiation is complex, and this complexity has obscured some simple
and general features of the light curves. In particular, {\em a reasonable
reproduction of the observed spectrum cannot guarantee a reasonable
representation of the thermal photon escape probability.}

Here we will use the 
simple model to illustrate some general points (a more accurate
discussion is in preparation with P. Nugent and P. Pinto). Note
that the escape operator for thermal photons might be better thought
of as diffusion down in energy space as well as out in radius; see
\cite{philron}. It will be an interesting challenge to develop
a better approximation to the thermal escape operator, which
is simple enough to allow analytic solutions.

In this analysis we will use only the solutions presented in \cite{arn82}
and earlier to emphasize the phemonena already contained in these early
efforts, which have only now had observational confirmation.

\section{The Shape of the Light Curve}

The bolometric
luminosity may be written as
\begin{equation}
 L = \epsilon_{Ni} M_{Ni} \Lambda(x,y).
\end{equation}
Here $ M_{Ni}$ is the mass of  $\rm^{56}Ni$;
$ \epsilon_{Ni}$ is the energy of radioactive decay of
 $\rm^{56}Ni$ per unit mass, divided
by the mean lifetime $ \tau_{Ni}$.
Actually  $\rm^{56}Co$ contributes as well.
This may be included in the $\Lambda$ function, with the
effect that the peak shifts higher in luminosity and later in time. 
In either case, $ \epsilon_{Ni} M_{Ni} $
is a convenient scale factor for the luminosity.

The shape of the light curve, which is independent of distance,
 is contained in $ \Lambda(x,y) $,
where 
\begin{equation}
 x = t / \tau_m,
\end{equation}
and
\begin{equation}
 y = \tau_m / 2 \tau_{Ni}.
\end{equation}
The effective escape time in a diffusing and expanding medium
is a logarithmic average of the expansion time $ \tau_{h}$
and the diffusion time $ \tau_{0}$,
\begin{equation}
\tau_m = \sqrt{ 2 \tau_{h} \tau_{0} },
\end{equation}
see \cite{arn82}. Note that $xy$ is simply time in units of
two mean lives of $\rm^{56}Ni$, or about 17.6 days. 
Although $y$ was defined within the particular analytic
context of simple diffusion in an expanding medium, it may be
more generally interpreted as a measure of the probability of
escape of energy by photon transport.

\begin{figure} 
  \vspace{16.5pc}
  \caption{ Lambda vs time (panel 1), normalized (panel 2).} 
\end{figure} 

The general character of these light curves is shown in figure~1, which
represent the solutions for \nuc{56}{Ni} decay alone.
For the observationally interesting case of maxima occurring around
$xy \approx 1$ as shown, we have
\begin{equation}
 \Lambda_{max} \approx 0.165/y,
\end{equation}
and
\begin{equation}
 t_{max} / 2 \tau_{Ni} \approx 0.42 + 0.48 y.
\end{equation}
These approximations apply to the simple case in which only the \nuc{56}{Ni}
heating is included, not that of \nuc{56}{Co} decay (the more general
case was calculated but not tabulated in \cite{arn82}). 
Near maximum light
($t \approx 2\tau_{Ni}$), the heating from \nuc{56}{Co} is equal to
that from \nuc{56}{Ni}, and dominates at later times. This additional
heating will increase both $t_{max}$ and $\Lambda_{max}$ relative
to the values give by these approximations, but make no qualitative
change.
The lower panel shows the effect of (1) renormalizing the
luminosity ($ \Lambda \rightarrow \Lambda/\Lambda_{max}$),
and (2) stretching the time scale to line up the maxima
($t \rightarrow t/t_{max}$ was used here). The curves lie
almost on top of each other, so that in this sense, the
shape is ``universal.'' 

For theoretical light curves the time of explosion is easily
defined, which is not true observationally. The observational
stretch includes a shift in time as well, 
$t  \rightarrow (t - t(0))/t_{max} $.

The light curves shown are those first presented by \cite{arn82}.
However, it was only after the discovery by \cite{perl97a} that
the observational data could be mapped into a universal curve
by a normalization of luminosity and a stretch of time scale
relative to the time of peak luminosity, that the analytic
solutions were plotted in this form. For this simple case, the
analytic solutions have this same property of (approximate)
universality as the data.

\section{The Early Light Curve}

Figure~1 also shows that all the light curves have a parabolic
dependence upon time during the earliest times after the explosion.
\cite{riess.near} present measurements of the earliest detections
of nearby SNIae, which delineate the rise behavior for 18 to 10 
days before maximum. 

According to \cite{riess.near}, Goldhaber has proposed a method of
determing the rise time of SNIae which is based upon the ``stretch''
method of \cite{perl97a}. \cite{riess.near,riess.far} have applied a similar
approach to the B-band light curves of a number of SNIae. Goldhaber proposed
that we describe the young SNIa as a homologously expanding fireball whose
luminosity is most sensitive to its increasing radius, rather than
effective temperature. The luminosity is then
\begin{equation}
L = \alpha ( t_{max} + t_r)^2,
\end{equation}
where $t_{max}$ is the time elapsed relative to maximum, $t_r$ is the
rise time, and $\alpha$ is the ``speed'' of the rise.

\begin{figure} 
  \vspace{16.5pc}
  \caption{ Scaled Luminosity of SNIae vs stretched time.} 
\end{figure} 
Figure~2 shows the data from \cite{riess.near}.
The squares represent their ten SNIae; upper limits have not been
plotted. The time coordinate has
been stretched by their stretch factors, and shifted so that $t=0$ corresponds
to the onset of the explosion. Thus, after stretching, the new time 
coordinate is $ t =  t_{max} + t_r$.
 Their $B$ magnitude has been rudely converted
to solar luminosities by simply ignoring bolometric corrections
(this is roughly correct, see \cite{riess.near}). In this linear plot,
the quadratic nature of the time dependence is obvious.

This behavior was predicted in 
 \cite{arn82}. Ten days before maximum is roughly ten days after
explosion, at which time the Co luminosity is about 0.3 of that
of Ni, and should not yet make a qualitative difference. 
Note that $xy = t/2\tau_{Ni}$,
so for early times ($t << \tau_{Ni}$), 
$\Lambda \approx x^2 = (t/2 \tau_{Ni})^2 /y^2$, so that
\begin{equation}
L = \epsilon_{Ni} M_{Ni} (t/2 \tau_{Ni}y)^2.
\end{equation}
The luminosity scale is set by the mass of \nuc{56}{Ni} and the shape
parameter $y$, the time dependence is quadratic in $t$, and
\begin{equation}
\alpha =  \epsilon_{Ni} M_{Ni} /( 2 \tau_{Ni} y )^2.
\end{equation}
This identifies Goldhaber's $\alpha$ with the solution parameters,
$2 \tau_{Ni}^2 y^2 = \kappa M/ 2 \beta c v_{sc},$
where  $\kappa$ is the effective opacity and $\beta \approx 13.7$ 
(see \cite{arn80}, Table~2), and $M_{Ni}$ which is the mass of \nuc{56}{Ni}.

The solid lines in figure~2 represent this solution for 
$(M_{Ni}/M_\odot)/y^2 = $ 0.25, 0.5 and 1.0. At early times
$(M_{Ni}/M_\odot)/y^2 \approx 0.4. $ For a popular estimate of
$(M_{Ni}/M_\odot) = 0.6$, we have $y =$ 1.2. As we shall see, this
is a plausible value.

\section{Higher Modes}
The behavior shown in figure~1 and figure~2 is based on
a theoretical model which assumes that the higher modes
in the spatial solution of the diffusion equation are small
(\cite{arn80}). These higher modes can be driven by 
\begin{enumerate}
\item a distribution of $\rm^{56}Ni$ which is different from the
distribution of energy in the fundamental mode for
diffusion,  \cite{philron},
\item a time dependence in the opacity (effective
escape parameter~$y$), or 
\item the interaction of
the exploding star with surrounding matter or a companion. 
\end{enumerate}
Such overtones modify
the shape of the light curve, and in principle can be detected
as a distance independent characteristic of SNIae.
The \cite{riess.near} data in figure~2 place limits on these effects.

\section{The Risetime}

Most theoretical models of SNIae predict significantly shorter risetimes
than are found (\cite{vaccaleib,riess.near}).
For example, \cite{hoef} give risetimes to visual maximum of 9 to 16 days,
with an average value of 14 days for single white dwarf explosions. 
The same difference is seen by \cite{philron}. \cite{riess.near} state:

``If these models are otherwise accurate, we concur with the conclusion of
Vacca \& Liebundgut (1996)
 that the model atmospheric opacity has been significantly
underestimated.
Past work suggests that deficient resonance line lists may be the culprit. 
By increasing the number of resonance lines from 500 to 100,000, the risetime
for models by Harkness (1991) increased by 8 days.''

By including the additional lines, Harkness decreased the escape probability
for thermal photons, which in our language means increasing $y$. This means
not only {\em atmospheric} opacity, but opacity at all depths. Because the
opacity is strongly frequency dependent (\cite{philron}), the atmosphere is
not a sharply defined radius. This makes the escape probability view a
clearer one. If maximum light occurs at 19.5 days, about 3.2 
half-lives, the \nuc{56}{Ni} has dropped to about 0.10 of its initial
value, and most of the decayed Ni is in the form of Co. The line lists
for Co and Ni are less extensive than for Fe, although these elements
have comparably complex atomic and ionic states. While this may not
affect spectral synthesis, which is often more sensitive to strong lines
(which are likely to be in even poor line lists), it would
affect the thermal escape probability. 

If the spectral synthesis modelling of SNIae is deficient in this way,
then attempts to use these models to infer global properties of SNIae
will inevitably be biased, and their use for detailed inferences concerning
distances and evolution of SNIae suspect. This argues for the relevance of the
simpler approach pioneered by Dave Branch (see \cite{nugent}), which
focuses on the atmosphere, and for the systematic effort to understand
the physics of the complex models in order to make them adequately robust
(\cite{baron,philron}). 

If we use a value of $y=1.2$ (see above), then we expect 
$\Lambda_{max} \approx 0.138$ and $t_{max} \approx 17.5$ days. This is
less that the value of 19.5 of \cite{riess.near}, but it is an
underestimate because heating from \nuc{56}{Co} decay will shift the
maximum luminosity to later times. The luminosity at maximum is then
$L/L_\odot = 32.0 \times 10^9 $, or a B magnitude of -18.35.
The addition Co luminosity brightens this by about -0.75 to -19.1.
This is to be compared to $M_B = -19.45$ of \cite{riess.near},
which is encouragingly consistent, given the crudeness of our
analysis.

\section{The Phillips Relation}

As is clear from the top panel in figure~1, the $\Lambda$
curves which peak earlier (the ones with smaller $y$),
have larger values at peak. Thus, to the extent that the 
nickel mass
$M_{Ni}$ does not change with $y$ (from event to event),
we have an {\em anti-Phillips} relations (\cite{phillips}).
How should $M(Ni)$ change with $y$?

Let us begin by examining a carbon-oxygen 
white dwarf of near Chandrasekhar
mass, which makes a transition to detonation after expanding
to some lower central density $\rho_{tran}$.
The material will be heated to a temperature which depends
primarily upon its current density, and the energy available
from burning. We may divide the star into layers, depending
upon whether its peak temperature allows explosive carbon
burning, oxygen burning, or silicon burning 
(see \cite{wac73,da96}).
We will ignore deviations of the mass-density structure from
that of an $n=3$ polytrope.
The transition to detonation may occur in a violent deflagration
or in the compressional phase following a mild deflagration.
Details may vary from the simple model
we use, depending upon the explosion mechanism and the 
progenitor characteristics.

\begin{figure} 
  \vspace{16.5pc}
  \caption{Polytropes n=3 yield vs central density at detonation.} 
\end{figure} 
The top panel of figure~3 shows the final composition expected
for a white dwarf, as a function of the central density it has
when it detonates.
At low density there is no burning, and the initial CO abundances
are preserved. Carbon burning produces mostly O and Mg at these
explosive temperatures. At still higher density, oxygen burning
makes Si, S, Ar, and Ca (SiCa). 
At the highest density, nickel is the dominant ash.

If all these changes in composition, throughout the white dwarf,
are converted into an implied explosion energy, we may relate
the explosion energy to the mass of nickel produced. This is 
shown in the lower panel. 
At the lowest densities for detonation, the burning does not
proceed through silicon burning to make Ni, but energy is
released from burning up to SiCa. There is an abrupt rise
to about 0.7 foe ($10^{51}$ ergs), and then a gradual increase
to over 1.3 foe, so
\begin{equation}
E_{SN}/(10^{51}{\rm erg} \approx 0.7 + (6/7)(M_{Ni}/M_\odot).
\end{equation}
This is related to the velocity scale through the distribution
of post-explosion velocity with density;
\begin{equation}
E_{SN}= {1\over 2}M<v^2>.
\end{equation}

If the composition has no effect on the opacity, or more precisely,
the thermal photon escape time, then the $y$ parameter depends upon
$M(Ni)$ through the velocity scale, or equivalently, the explosion
energy. Thus,
\begin{equation}
y \propto \sqrt{\kappa M/v_{sc}} \rightarrow M^{3/4}/E_{SN}^{1/4},
\end{equation}
which is a relatively weak dependence on $E_{SN}$.

Collecting results,
\begin{eqnarray}
L& = & \epsilon_{Ni} M_{Ni} \Lambda_{max} \cr
 & \propto & M_{Ni}/y \cr
 & \propto & y^3
\end{eqnarray}
where the last result assumes $ E>>0.7$ foe. This gives the observed sense
of the Phillips relation: brighter SN have broader light curves. 
However, there is a potential difficulty here (\cite{philron}): if the
\nuc{56}{Ni} is distributed as a central sphere of pure Ni, then the
average distance to the surface increases with $M_{Ni}$, and so does
the probability of leakage from gamma and x-ray escape. This changes
the light curve shape, and may destroy the sense of the Phillips relation.
Alternatively, if the Ni distrubution is not a strong function of $M_{Ni}$
at a time several days after the explosion, the previous result holds. 
The production of a Phillips relation that agrees with observation depends
upon the nature of the explosion assumed, but may not be difficult to get
for some simple and attractive models.

\section{Conclusions}
The new data allow us to independently determine the parameters in
the analytic models of SNIae. The predictions for the premaximum
behavior of the light curves are confirmed, and 
it will be possible to plance new contraints on the nature of 
the progenitors and explosions. 
Even at the crude level sketched here,
it is possible to get self consistent values, and a more accurate
computation following this logic is warranted. In contrast, the
radiative transfer models of \cite{hoef} fail to give the correct
rise time, probably due to incomplete line lists, and are likely
to give biased results when applied to the description of such
global properties of SNIae, at least until this weakness is corrected. 

\begin{acknowledgments}

Enlightening discussions with participants in this workshop and
the Aspen Center for Physics workshop on Type~Ia Supernovae
are gratefully acknowledged.  In particular, Pat Nugen, Phil Pinto,
Paulo Mazzali, 
Adam Riess, Eddy Baron, Brian Schmidt, and Saul Perlmutter made
specific comments which affected the development above.
  This work was supported 
in part by DOE grant DE-FG03-98DP00214/A001.

\end{acknowledgments}


\begin{thebibliography}{} 


\bibitem[Arnett (1980)]{arn80} 
     {\sc Arnett, D.}, 1980, {\it Ap. J.}~{\bf 237}, 541.

\bibitem[Arnett (1982)]{arn82} 
     {\sc Arnett, D.}, 1982, {\it Ap. J.}~{\bf 253}, 785.

\bibitem[Arnett (1996)]{da96} 
     {\sc Arnett, David} 1996, {\it Supernovae and Nucleosynthesis\/},
     Princeton University Press. 

\bibitem[Baron, \etal\ (1999)]{baron}
     {\sc Baron, e., Branch, D., Hauschildt, P. H., Filippenko, A. V.,
     \& Kirshner, R. P.}, 1999, {\it Ap. J.}, in press

\bibitem[Hamuy, \etal\ (1995)]{calan}
     {\sc Hamuy, M., Phillips, M. M., Maza, J.,
     Suntzeff, N. B., Schommer, R. A., \& Aviles R.}, 1995, 
     {\it A. J.}~{\bf 109}, 1

\bibitem[Harkness (1991)]{hark}
     {\sc Harkness, R. P.}, 1991, in {\it SN1987A and Other Supernovae},
     ed. I. J. Danziger \& K. Kjar (Garching ESO).

\bibitem[H\"oflich \& Khokhlov (1996)]{hoef}
     {\sc H\"oflich, P., \& Khokhlov, A.}, 1996,
     {\it Ap. J.}~{\bf 457}, 500.

\bibitem[Nugent, \etal\ (1995)]{nugent}
     {\sc Nugent, P., Branch, D., Baron, E., Figher, A., \& Vaughan, T.},
     1995, {\it Phys. Rev. Lett.}~{\bf 75}, 394.

\bibitem[Perlmutter \etal\ (1997a)]{perl97a}
     {\sc Permutter, S., \etal\ }, 1997a, in {\it Thermonuclear
     Supernovae\/}, ed., Ruiz-Lapuente, P., Isern, J., \& Canal, R.,
     p. 749, Kluwer Publishers, Dordrecht.


\bibitem[Perlmutter,  \etal\ (1997b)]{perl97b}
     {\sc Permutter, S. \etal\ }, 1997b
     {\it Ap. J.}~{\bf 483}, 565.

\bibitem[Perlmutter, \etal\ (1999a)]{perl99a}
     {\sc Permutter, S. \etal\ }, 1999a
     {\it Ap. J.}~{\bf 517}, 565.

\bibitem[Perlmutter (1999b)]{perl.this}
     {\sc Permutter, S. } 1999b, this conference.

\bibitem[Phillips (1993)]{phillips}
     {\sc Phillips, M. M.} 1993, {\it Ap. J.}~{\bf 413}, L105.

\bibitem[Pinto \& Eastman (1999)]{philron} 
     {\sc Pinto, P. A., \& Eastman, R.},
     1999, {\it Ap. J.},~ in press.

\bibitem[Riess, Press \& Kirshner (1995)]{rpk}
     {\sc Riess, A. G., Press, W. H., \& Kirshner, R. P.}, 1995,
     {\it Ap. J.}~{\bf 438}, L17.

\bibitem[Riess,  \etal\ (1999a)]{riess.near}
     {\sc Riess, A. G., Filippenko, A. V., Li, W., Treffers, R. R.,
     Schmidt, B. P., Qui, Y., Hu, J., Armstrong, M., Faranda, C.,
     \& Thouvenot, E.}, 1999a, {\it Ap. J.}, in press.

\bibitem[Riess, \etal\ (1999b)]{riess.far}
     {\sc Riess, A. G., Filippenko, A. V., Li, W., \& Schmidt, B. P.},
     1999b, {\it Ap. J.}, in press.

\bibitem[Schmidt, \etal\  (1999)]{hiz}
     {\sc Schmidt, B., \etal\ }, 1998
     {\it Ap. J.}~{\bf 5507}, 46.

\bibitem[Vacca \& Leibundgut (1996)]{vaccaleib}
     {\sc Vacca, W. D., \& Leibundgut, B.}, 1996, 
     {\it Ap. J.}~{\bf 471}, L37.

\bibitem[Woosley, Arnett, \& Clayton 1973]{wac73} 
     {\sc Woosley, S. E., Arnett, W. D., \& Clayton, D. D.},
     1973, {\it Ap. J. Suppl.}~{\bf 26}, 231.

\end{thebibliography}
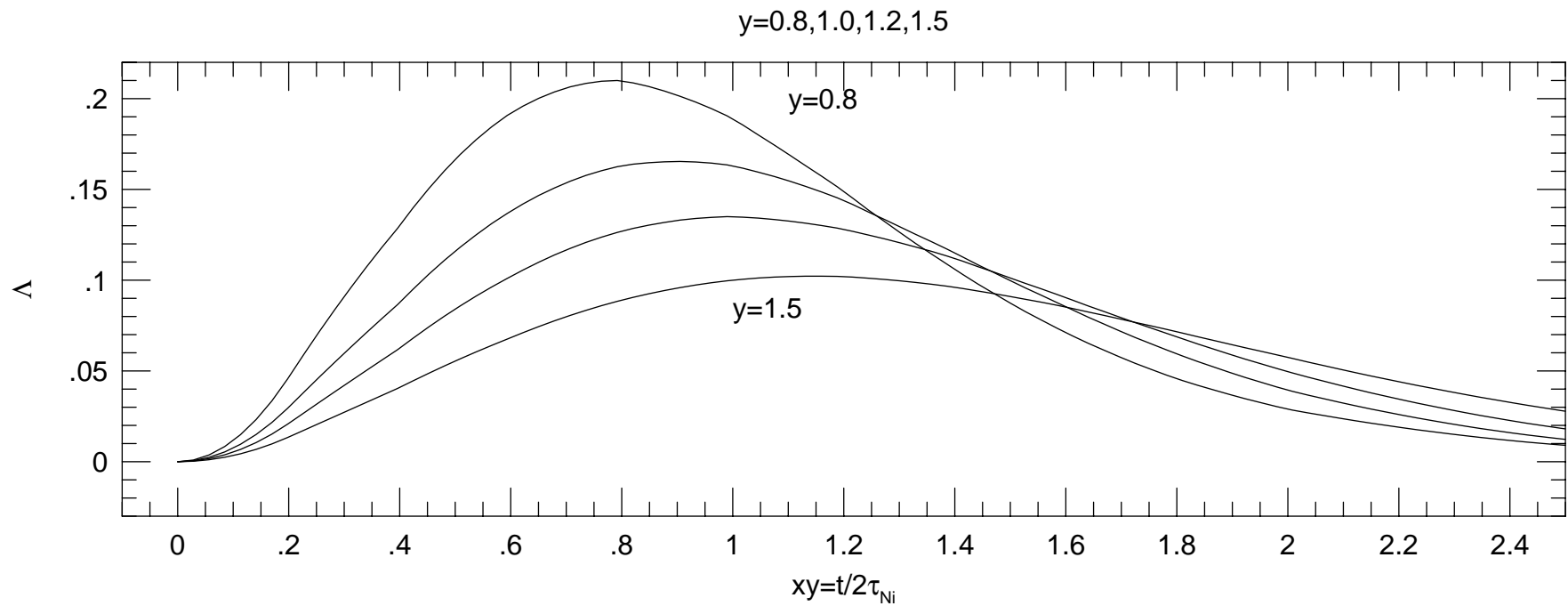
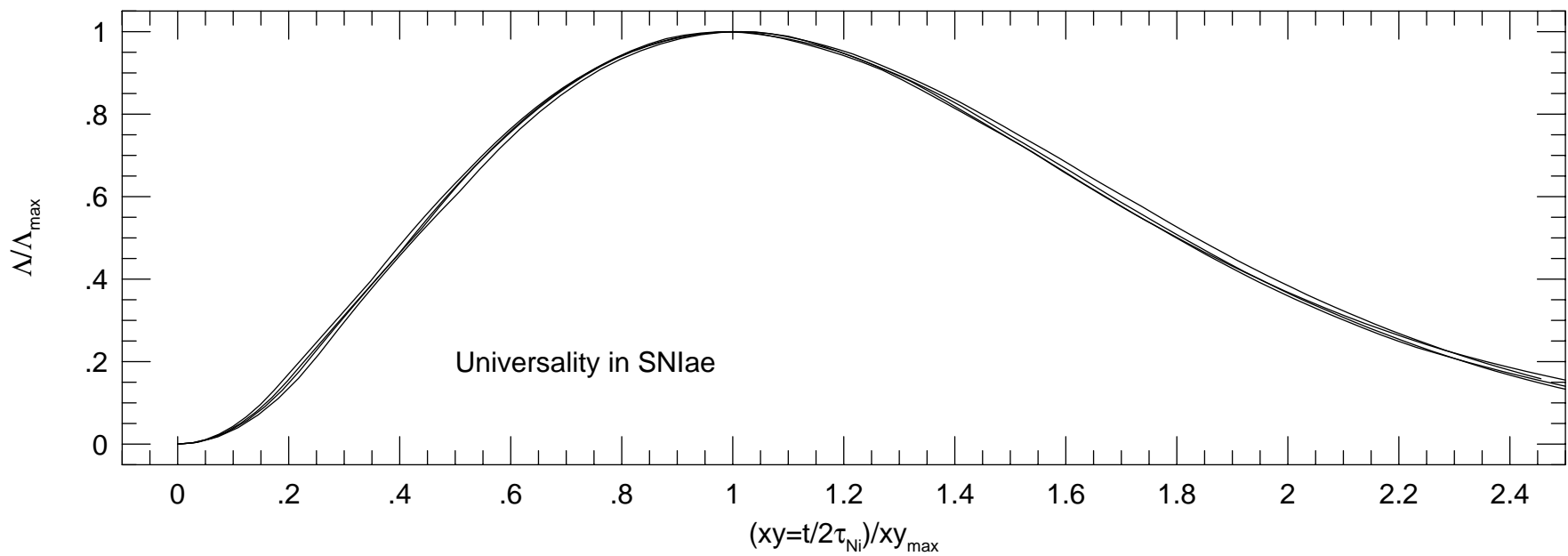
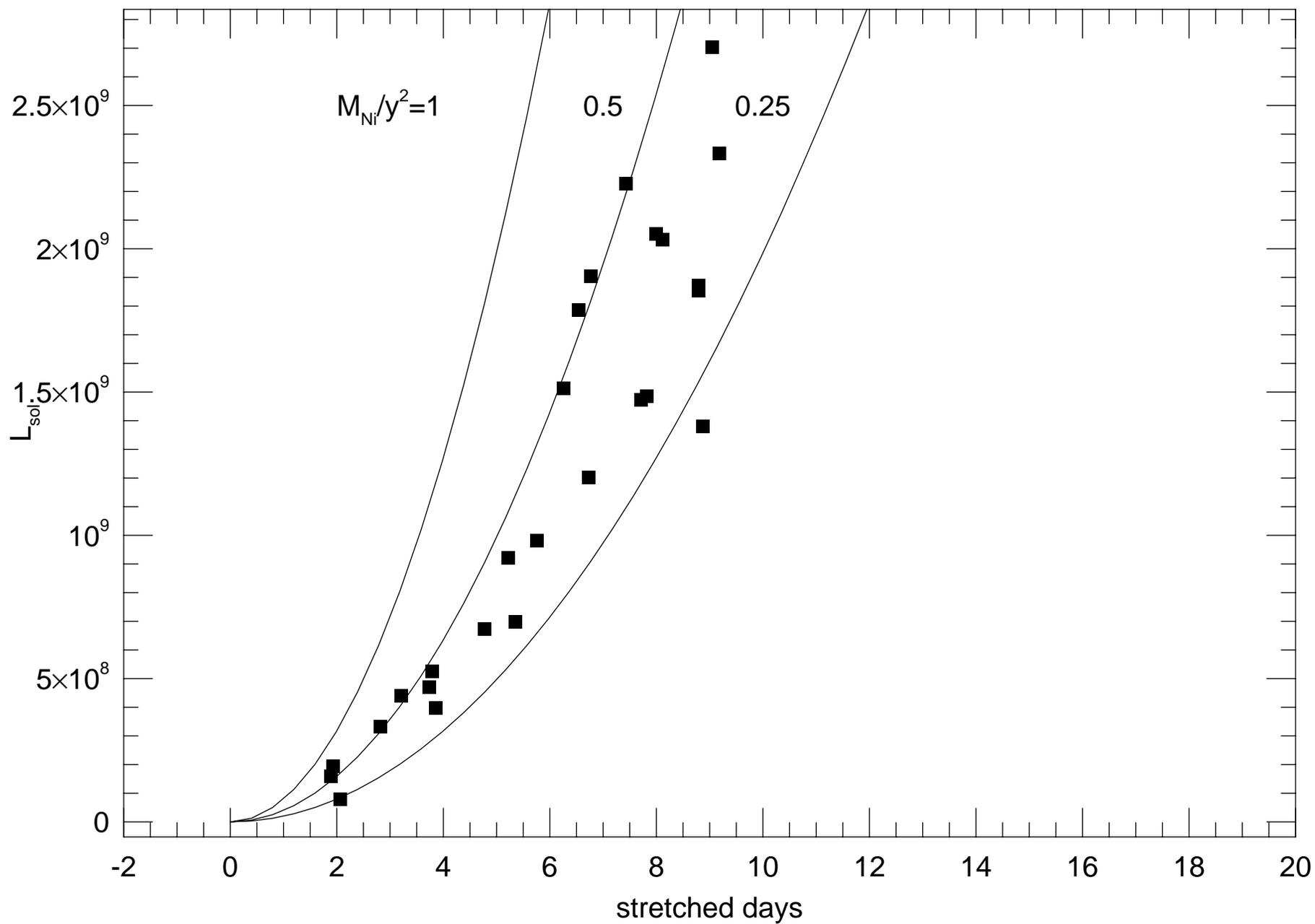
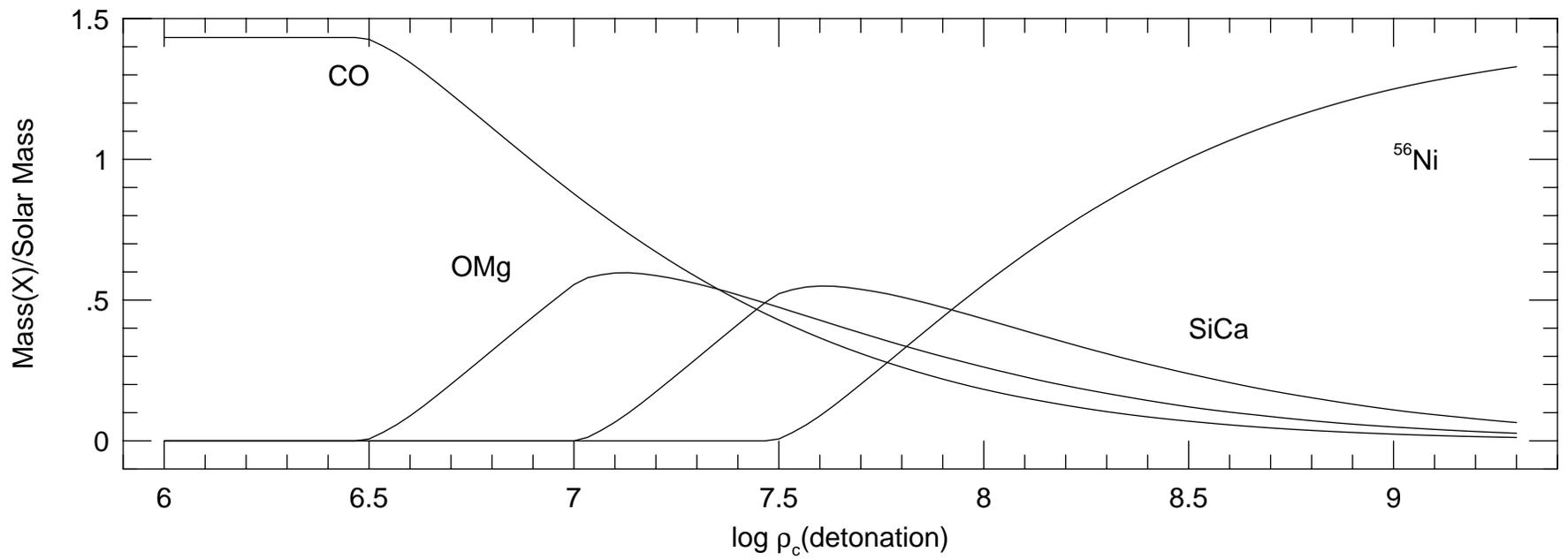
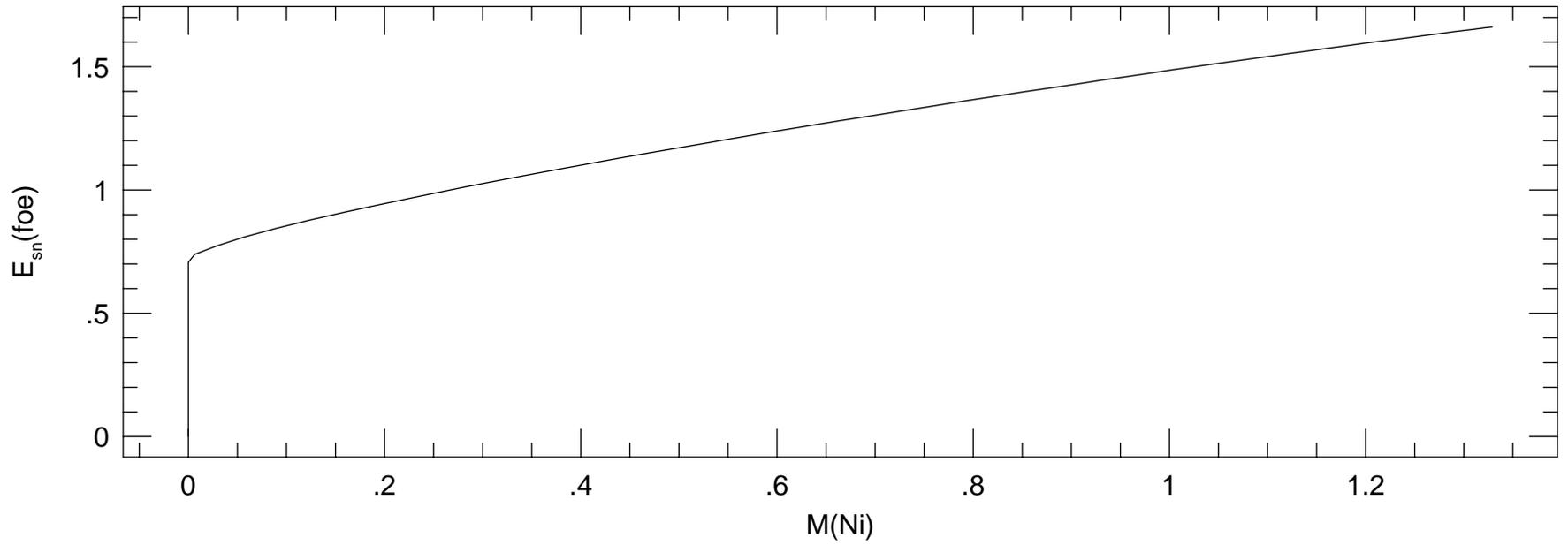
\end{document}